\documentclass[11pt, a4paper]{article}

\usepackage[a4paper, total={16cm, 25cm}]{geometry}
\usepackage{amsmath}
\usepackage{graphicx}
\usepackage{natbib}
\usepackage{amsthm}
\usepackage{amsfonts}
\usepackage{color}
\usepackage{multirow}
\usepackage{tikz}

\newtheorem{rational for conjecture}{Rational for Conjecture}

\DeclareMathOperator*{\argmin}{arg\,min}

\begin{document}

\LARGE

\begin{center}
\noindent \textbf{Efficient Quantile Tracking Using an Oracle}
\end{center}

\large

\begin{center}
Hugo L. Hammer\footnote{Oslo Metropolitan University and Simula Metropolitan Center} \footnote{Corresponding author. Email: \texttt{hugo.hammer@oslomet.no}}, Anis Yazidi\footnote{Oslo Metropolitan University}, Michael A. Riegler\footnote{Simula Metropolitan Center} and H{\aa}vard Rue\footnote{King Abdullah University of Science \& Technology}
\end{center}

\normalsize

\begin{abstract}

For incremental quantile estimators the step size and possibly other tuning parameters must be carefully set. However, little attention has been given on how to set these values in an online manner. In this article we suggest two novel procedures that address this issue. 

The core part of the procedures is to estimate the current tracking mean squared error (MSE). The MSE is decomposed in tracking variance and bias and novel and efficient procedures to estimate these quantities are presented. It is shown that estimation bias can be tracked by associating it with the portion of observations below the quantile estimates. 

The first procedure runs an ensemble of $L$ quantile estimators for wide range of values of the tuning parameters and typically around $L = 100$. In each iteration an \textit{oracle} selects the best estimate by the guidance of the estimated MSEs. The second method only runs an ensemble of $L = 3$ estimators and thus the values of the tuning parameters need from time to time to be adjusted for the running estimators. The procedures have a low memory foot print of $8L$ and a computational complexity of $8L$ per iteration.

The experiments show that the procedures are highly efficient and track quantiles with an error close to the theoretical optimum. The \textit{Oracle} approach performs best, but comes with higher computational cost. The procedures were further applied to a massive real-life data stream of tweets and proofed real world applicability of them.
\end{abstract}

\section{Introduction}

The volumes of automatically generated data are constantly increasing \citep{ramirez2017survey} with more urgent demand for being analyzed in real-time \citep{kejariwal2015real}. Conventional statistical and data mining techniques are constructed for offline situations and therefore are often not applicable for such real-time analysis \citep{krempl2014open}. Thus a wide range of streaming algorithms are continuously being developed by the data mining community addressing a range of real-time tasks such as clustering, filtering, cardinality estimation, estimation of moments or quantiles, predictions, dimensionality reduction and anomaly detection \citep{kejariwal2015real}.

An inherent part of almost any real-time method is how rapid the history of the data stream should be forgotten. The forgetting mechanism should be adjusted depending on the properties of the data stream which again can vary over time. A prominent example to address this issue is the adaptive windowing approach (ADWIN) \citep{bifet2007learning} where a historic and a current window of observations are compared to look for changes in the properties of the data stream. The sizes of the windows are further adapted, and properties of the data stream can be computed from the adaptive windows. \citet{costa2007adaptive} suggests methods to adjust the step sizes of stochastic approximation algorithms related to the frequency of regime switches. There is also a wide range of event detection based methods that rapidly adjust the current estimate when an event is detected, see e.g. \citet{ross2014sequential,hammer2018parameter}. 

In this paper we address the issue when the objective is to track quantiles of data streams. To the best of our knowledge this is the first article addressing this. Quantiles are useful to characterize data stream distribution in a flexible and non-parametric way \citep{luo2016quantiles} and have been used for a wide range of applications such as portfolio risk measurement in the stock market \citep{gilli2006application,abbasi2013bootstrap}, fraud detection \citep{zhang2008detecting}, signal processing and filtering \citep{stahl2000quantile}, climate change monitoring \citep{zhang2011indices}, SLA violation monitoring \citep{sommers2007accurate,sommers2010multiobjective}, network  monitoring \citep{choi2007quantile,liu2018accurate}, Monte Carlo simulation \citep{wang2016quantiles}, structural health monitoring \citep{gregory2018quantile}, non-parametric statistical testing \citep{lall2015data}, concept drift detection \citep{hammer2019new} and Tukey depth estimation \citep{hammer2020estimating}.

We focus on a family of lightweight and efficient methods called incremental quantile estimators, see \citet{yazidi2017multiplicative,hammer2019new} for recent reviews. The methods are based on performing small updates of the quantile estimate every time a new sample is received from the data stream and the methods can document state-of-the-art performance on quantile tracking. We suggest two new procedures that select and tune the update size, respectively, by the guidance of an estimate of the current tracking mean squared error (MSE). Naturally the MSE cannot be computed directly since the true quantile is unknown. The MSE is estimated by decomposing it in estimation variance and bias and novel and efficient procedures to estimate these quantities are presented. In particular we show that estimation bias can be associated with the portion of observations below the quantile estimates. The procedures are highly efficient with $O(1)$ memory requirement and $O(1)$ computational complexity per iteration. 

The paper is organized as follows. Section \ref{sec:incr} gives a short presentation of incremental quantile estimators, Section \ref{sec:MSE} presents how to estimate the current tracking MSE and how to use it to adapt the quantile estimates. Sections \ref{sec:synt} and \ref{sec:real} evaluate the suggested procedure using synthetic and real-life data streams while Section \ref{sec:closrem} gives some closing remarks.

\section{Incremental Quantile Estimators}
\label{sec:incr}

Let $X_t \sim f_t(x)$ represent possible outcomes from a data stream at time $t$, $x_t$ a random sample and $Q_{t,q}$ the quantile associated with probability $q$, i.e $P(X_t \leq Q_{t,q}) = F_t(Q_{t,q}) = q$.

Incremental quantile estimator update a quantile estimate every time a new observation is received. The algorithms are initiated with an estimate $\widehat{Q}_{0,q}(\lambda)$  and further recursively updated 
\begin{align}
  \label{eq:6}
  \begin{split}    
  \widehat{Q}_{t+1,q}(\lambda) &\leftarrow \widehat{Q}_{t,q}(\lambda) + \lambda D_1\left(q, \widehat{Q}_{t,q}(\lambda)\right) \hspace{5mm} \text{ if } x_t \geq \widehat{Q}_{t,q}(\lambda) \\
  \widehat{Q}_{t+1,q}(\lambda) &\leftarrow \widehat{Q}_{t,q}(\lambda) - \lambda D_2\left(q, \widehat{Q}_{t,q}(\lambda)\right) \hspace{5mm} \text{ if } x_t < \widehat{Q}_{t,q}(\lambda) 
  \end{split}
\end{align}
where $D_1$ and $D_2$ are positive and can be deterministic or random. The estimator is intuitive in the sense that if the received observation is above (below) the current estimate, the estimate is increased (decreased). The functions are typically further constructed to ensure that the estimator converges to the underlying true quantile \citep{Tierney1983}. A prominent example is the deterministic based multiplicative incremental quantile estimator (DUMIQE) \citep{yazidi2017multiplicative} where $D_1(q, \widehat{Q}_{t,q}(\lambda)) = q \widehat{Q}_{t,q}(\lambda)$ an $D_2(q, \widehat{Q}_{t,q}(\lambda)) = (1 - q) \widehat{Q}_{t,q}(\lambda)$. Another example is the Frugal estimator \citep{ma2013frugal} where $D_1(q, \widehat{Q}_{t,q}(\lambda)) = I(q < U)$ and $D_2(q, \widehat{Q}_{t,q}(\lambda)) = I(1-q < U)$ and $U$ uniformly distributed on $[0,1]$ and $I(\cdot)$ the indicator function.

The tuning parameter $\lambda$ determines update size. If the data stream distribution changes rapidly (slowly) with time, a high (small) should be used. Further, the step size should be adjusted to the scale of the data. 

\section{Adaptive Updating of Tuning Parameters}
\label{sec:MSE}

In this section we explain how to estimate the current quantile tracking mean squared error (MSE) and how to use this to efficiently track the true quantile. The MSE estimator uses the well-known bias and variance decomposition
\begin{equation}
  \label{eq:1}
  \text{MSE}_t = \mu\left(\widehat{Q}_{t,q} - Q_{t,q} \right)^2 = \underbrace{\left(\mu(\widehat{Q}_{t,q})- Q_{t,q} \right)^2}_{\text{Bias}^2} + \underbrace{\sigma^2\left(\widehat{Q}_{t,q} \right)}_{\text{Variance}}
\end{equation}
where $\mu(\cdot)$ and $\sigma^2(\cdot)$ refer to expectation and variance, respectively. For notational convenience, references to $\lambda$ are avoided in this part.

The variance in Eq. \eqref{eq:1} is tracked using exponential smoothing, but with some careful considerations. For dynamically changing data streams, $\sigma^2(\widehat{Q}_{t,q})$ typically is very small relative to the dynamic changes of the data stream\footnote{see Figure \ref{fig:1} for an example.} making this a hard tracking problem. A common approach is to rely on the result that $\sigma^2(\widehat{Q}_{t,q}) = \mu(\widehat{Q}_{t,q}^2) - \mu(\widehat{Q}_{t,q})^2$ and separately track the two expectations, but this approach is prone to loss of precision and cannot guarantee that the estimate will be positive. Instead we suggest the following estimator 
\begin{align}
  \label{eq:2}
  \widehat{\mu}\left(\widehat{Q}_{t,q}\right) &\leftarrow (1 - \alpha) \hat{\mu}\left(\widehat{Q}_{t-1,q}\right) + \alpha \widehat{Q}_{t,q}\\
  \label{eq:5} \widehat{\sigma^2}\left(\widehat{Q}_{t,q}\right) &\leftarrow (1 - \beta) \widehat{\sigma^2}\left(\widehat{Q}_{t-1,q}\right) + \beta \left(\widehat{Q}_{t,q} - \widehat{\mu}\left(\widehat{Q}_{t,q}\right)\right)\left(\widehat{Q}_{t,q} - \widehat{\mu}\left(\widehat{Q}_{t-1,q}\right)\right)
\end{align}
which is a generalization of \cite{knuth2014art} that used $\beta = \alpha$. When $\alpha \neq \beta$ the estimator will be biased, but our experiments showed that the bias is small even for large difference between $\alpha$ and $\beta$. If a too small value of $\alpha$ is used in Eq. \eqref{eq:2}, the estimator will not be able the track the expectation resulting in serious overestimation of the variance in Eq. \eqref{eq:5}. By rather using a large value, say $\alpha = 0.5$, combined with a small value of $\beta$ to efficiently smooth the volatile local variance estimates worked well in our experiments. 

Naturally the Bias$^2$ term in Eq. \eqref{eq:1} cannot be computed since $Q_{t,q}$ is unknown. However, the following computation shows that it can be associated with the \textit{expected portion of observations below the quantile estimate} which is possible to estimate:
\begin{align}
  \text{Bias}^2 &= \left(\mu(\widehat{Q}_{t,q})- Q_{t,q} \right)^2\\
  &= \left(\mu(G_t(\widehat{q}_t))- G_{t,q} \right)^2 \hspace{1.5cm} \text{where } G_t = F_t^{-1}\\
                                     \label{eq:3} &\approx \left(\mu\left(G_{t,q} + G'_{t,q}(\widehat{q}_t - q)\right) - G_{t,q}\right)^2\\
                                                  &= \left(G_{t,q} + G'_{t,q}(\mu\left(\widehat{q}_t\right) - q) - G_{t,q}\right)^2\\
  \label{eq:7} &= (G'_{t,q})^2(\mu\left(\widehat{q}_t\right) - q)^2
\end{align}
where Eq. \eqref{eq:3} is derived using a first order Taylor approximation. $H_q\left(\mu\left(\widehat{q}_t\right)\right) = \left(\mu\left(\widehat{q}_t\right) - q\right)^2$ in Eq. \eqref{eq:7} is tracked in a two stage procedure similar to the variance procedure above
\begin{align}
  \label{eq:4}
  \widehat{\mu}\left(\widehat{q}_t\right) &\leftarrow (1 - \gamma) \widehat{\mu}\left(\widehat{q}_{t-1} \right) + \gamma I\left(x_t \leq \widehat{Q}_{t,q}\right) \\
  \widehat{H}_q\left( \widehat{\mu}\left(\widehat{q}_t\right)\right) &\leftarrow (1 - \kappa) \widehat{H}_q\left( \widehat{\mu}\left(\widehat{q}_{t-1} \right)\right) + \kappa \left(\widehat{\mu}\left(\widehat{q}_t\right) - q\right)^2
\end{align}
Finally $G'_{t,q}$ is estimated by tracking an auxiliary quantile for a second probability $\widetilde{q}$ close to $q$ using the same value of $\lambda$
\begin{align}
  \widehat{G}'_{t,q} \leftarrow (1 - \eta) \widehat{G}'_{t-1,q} + \eta \frac{\widehat{Q}_{t,q} - \widehat{Q}_{t,\widetilde{q}}}{q - \widetilde{q}}
\end{align}
An estimate of the MSE is now obtained by substituting the estimates into \eqref{eq:1}
\begin{equation}
  \label{eq:15}
  \widehat{\text{MSE}}_t = \left(\widehat{G}'_{t,q}\right)^2 \widehat{H}_q\left( \widehat{\mu}\left(\widehat{q}_t\right)\right) + \widehat{\sigma^2}\left(\widehat{Q}_{t,q}\right)
\end{equation}
A useful property of the MSE estimator is that it does not use the values of a data stream directly making it \textit{robust} to outliers if the quantile estimator is robust to outliers which is the case for most incremental estimators. 

\begin{figure}
  \centering
  \begin{tikzpicture}[scale=1, transform shape]
\tikzstyle{every node} = [rectangle, fill=gray!30]
\node (oracle) at (3.5,2.5) {\textbf{Oracle} $-$ Current estimate: $\widehat{Q}_{t,q}(\widetilde{\lambda}_t)$, where $\widetilde{\lambda}_t = \argmin_{\lambda_l, l \in 1,\ldots,L} \widehat{\text{MSE}}_t(\lambda_l)$};
\node (MSE1) at (-2,1) {$\widehat{\text{MSE}}_{t}(\lambda_1)$};
\node (Q1) at (-2,-2) {\includegraphics[width=0.25\textwidth]{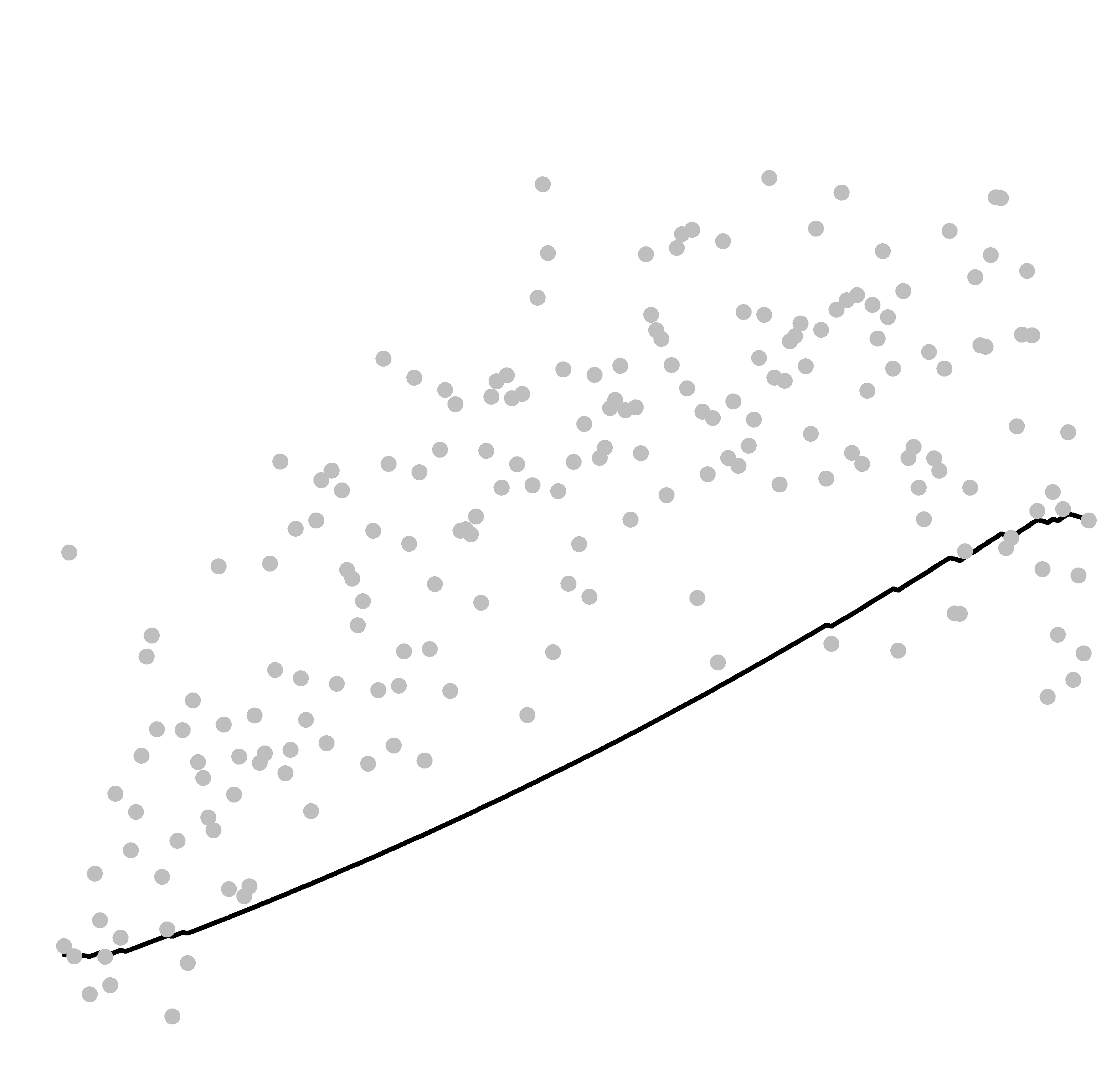}};
\draw[->,thick] (MSE1.north) -- (oracle.south);
\draw[->,thick] (Q1.north) -- (MSE1.south);
\node (MSE2) at (3,1) {$\widehat{\text{MSE}}_{t}(\lambda_2)$};
\node (Q2) at (3,-2) {\includegraphics[width=0.25\textwidth]{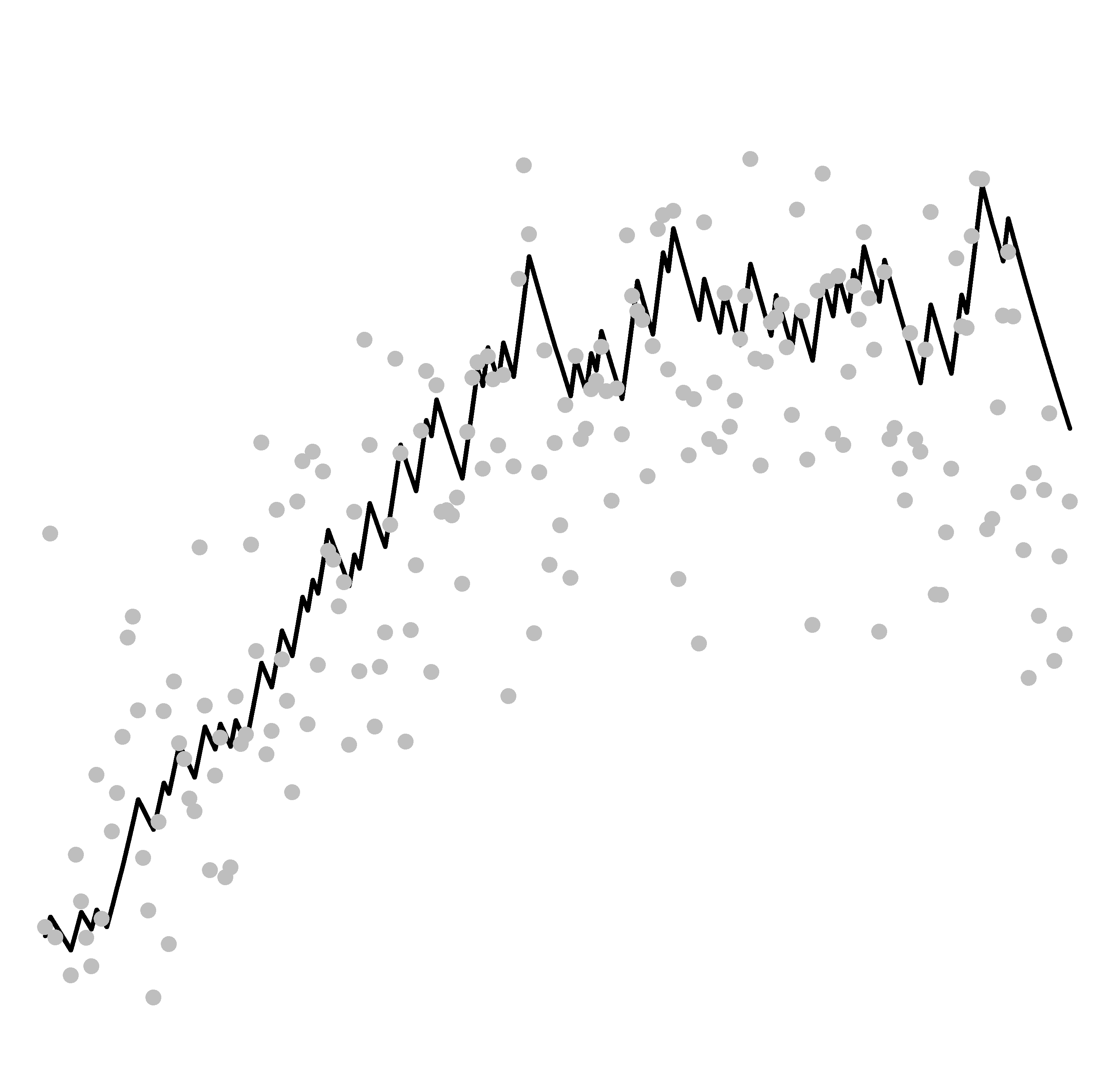}};
\draw[->,thick] (MSE2.north) -- (oracle.south);
\draw[->,thick] (Q2.north) -- (MSE2.south);
\node (dots) at (6, 1) {$\cdots$};
\node (dots) at (6, -2) {$\cdots$};
\node (MSEL) at (9,1) {$\widehat{\text{MSE}}_{t}(\lambda_L)$};
\node (QL) at (9,-2) {\includegraphics[width=0.25\textwidth]{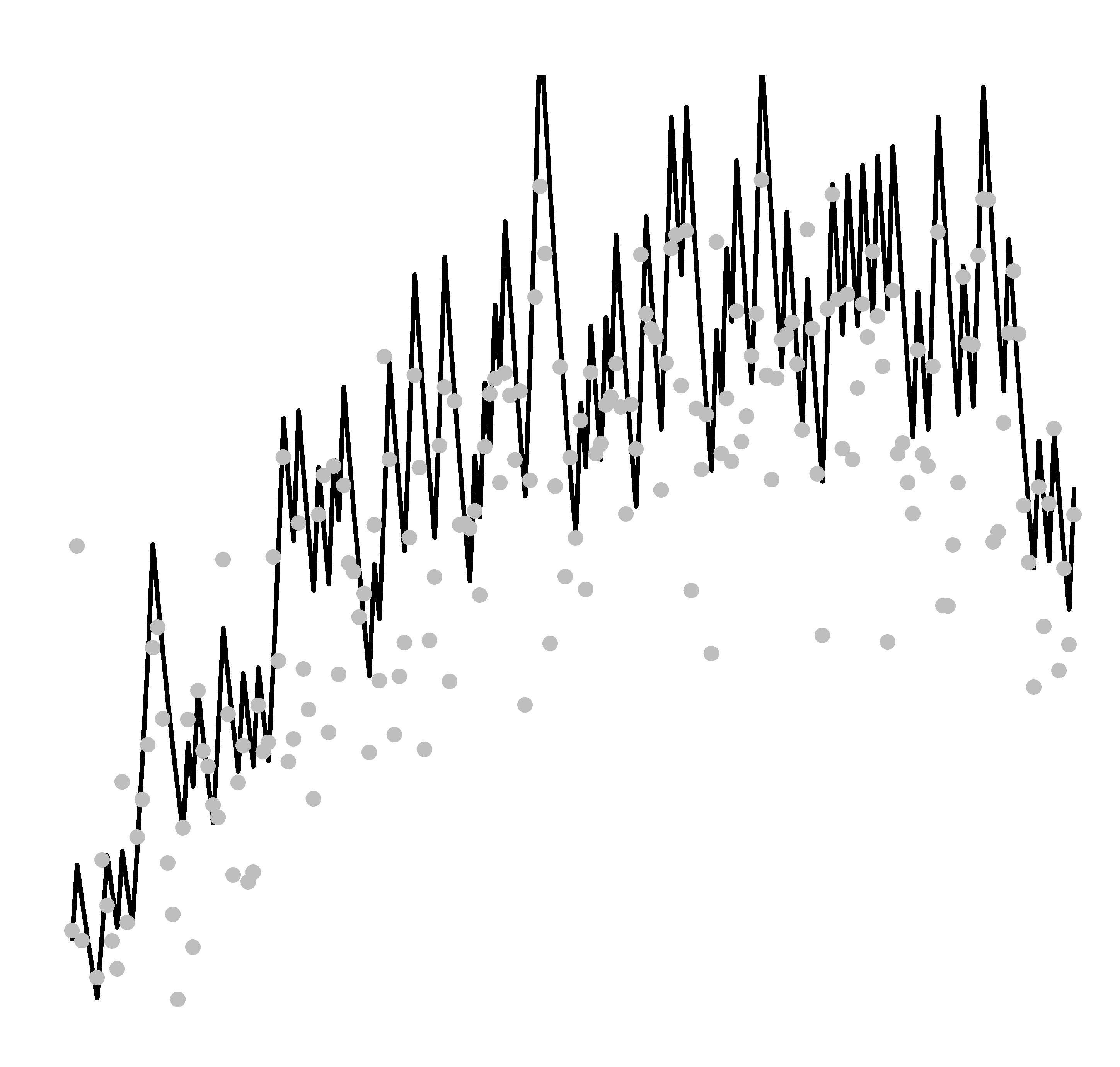}};
\draw[->,thick] (MSEL.north) -- (oracle.south);
\draw[->,thick] (QL.north) -- (MSEL.south);
\end{tikzpicture}
\caption{An overview of the \textit{Oracle} approach. In the bottom row, the gray dots show observations from the data stream. The black curves show tracking of the $q = 0.7$ quantile for different values of $\lambda$ ($\lambda_1$, $\lambda_2$ and $\lambda_L$). The information in the panels are used to estimate MSE for each value of $\lambda$ using the procedure in Section \ref{sec:MSE}. The MSE estimates are sent to the Oracle which selects the current quantile estimate.}
  \label{fig:5}
\end{figure}
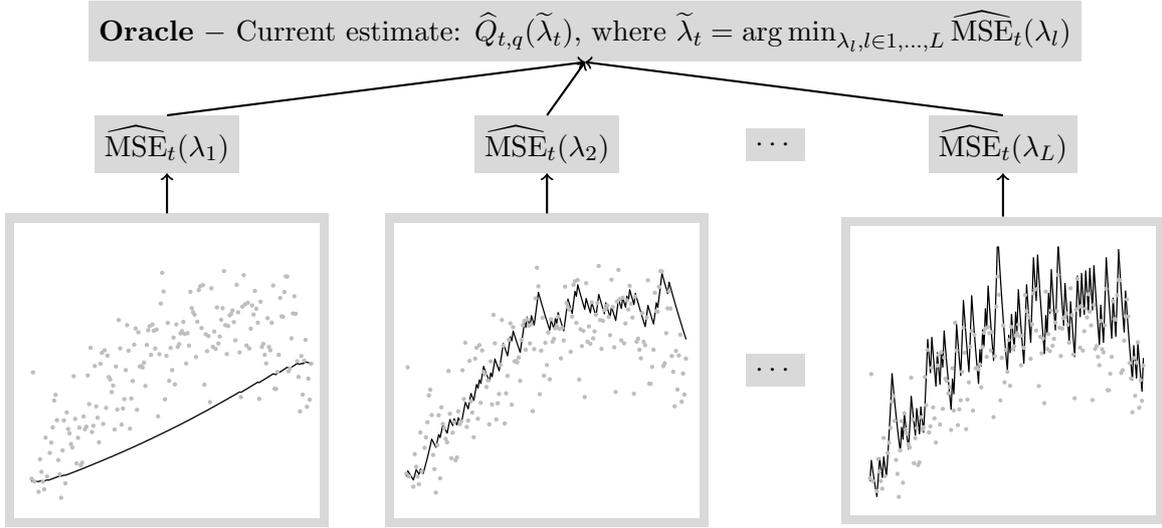

We suggest two strategies to efficiently track the true quantile using Eq. \eqref{eq:6}. \\
\textbf{Oracle}: Let $\lambda_1 < \lambda_2 < \cdots < \lambda_L$ span all reasonable values for the tuning parameter and track $Q_{t,q}(\lambda_l)$, $l = 1,\ldots,L$ using Equations \eqref{eq:6}. For each $Q_{t,q}(\lambda_l)$ estimate the associated MSEs, $\widehat{\text{MSE}}_t(\lambda_l)$, $l = 1,\ldots,L$, using \eqref{eq:15}. Let $\widetilde{\lambda}_t = \argmin_{\lambda_l, l \in 1,\ldots,L} \widehat{\text{MSE}}_t(\lambda_l)$ and let the current quantile estimate be given by $\widehat{Q}_{t,q}(\widetilde{\lambda}_t)$. Naturally, we may get large fluctuations in the values of $\lambda$ used and one may limit updates only to neighbouring values (friction). We denote this approach the \textit{Oracle} approach, since we can imagine an oracle that monitors the individual quantile tracking procedures and use the estimated MSEs to select the best current quantile estimate, without disturbing the quantile or MSE tracking procedures. The procedure is illustrated in Figure \ref{fig:5} where the objective is to track the $q = 0.7$ quantile of the data stream (gray dots). In the bottom left panel the quantile is tracked with a small update size, $\lambda_1$, resulting in a high estimation bias according to Eq. \eqref{eq:7} (the portion of observations below the quantile estimates is far below the target $q = 0.7$). Further in the bottom right panel the quantile is tracked with a large update size, $\lambda_L$, resulting in high tracking variance according to Eq. \eqref{eq:5}. The estimator in the middle panel seem to have both fairly small bias and variance resulting in a small MSE.

A potential challenge with the approach, is that with no knowledge about the data stream it may be difficult to select the range of $\lambda$'s. A potential solution is that if the oracle selects the quantile estimate for a $\lambda$ close to $\lambda_1$ (or $\lambda_L$) additional estimators with values of $\lambda$ below $\lambda_1$ (or above $\lambda_L$) are included. The approach needs to store a total of $8L$ values and perform a total of $8L$ operations for each received sample from the data stream where $L = 100$ is a representative value. Thus for extremely massive data streams, this may pose computational challenge. An example is when incremental quantile estimators are used to track depth contours requiring to track thousands or even millions of quantile estimates \citep{hammer2020estimating}. \\
\textbf{HIL}: A computationally less demanding approach is to only track the quantile for a high, intermediate and low (HIL) value of $\lambda$. Thus, the procedure only requires to store a total of $8\cdot 3$ values and perform a total of $8\cdot 3$ operations for each received data stream sample.
\begin{itemize}
\item Run three quantile tracking algorithms in parallel using tuning parameters $\lambda_{\text{1ow}} = \lambda/a$, $\lambda_{\text{intermediate}} = \lambda$ and $\lambda_{\text{high}} = a\lambda$, $a > 1$ and track the MSE for each of them.
\item Every $M$ iterations, update $\lambda$:
  \begin{itemize}
  \item If MSE is smallest for $\lambda_{\text{1ow}}$ (or $\lambda_{\text{high}}$), reduce (or increase) the value of the tuning parameter for the three estimators by setting $\lambda \leftarrow \lambda/a$ (or $\lambda \leftarrow a\lambda$). Restart the three quantile estimators initialized with the currently best quantile estimate, i.e. for $\lambda_{\text{1ow}}$ (or $\lambda_{\text{high}}$).
   \item If MSE is smallest using $\lambda_{\text{intermediate}}$, no updates are done.
  \end{itemize}  
\end{itemize}
In practice changes in data stream dynamics happen slowly or rarely (else it would be part of the original dynamics) and it makes sense to use a high value of $M$, say $1,000$. Further the values of the tuning parameters $\beta, \gamma, \kappa$ and $\eta$ should be chosen such that when an update of $\lambda$ is performed, the estimate of the MSE has converged and at the same time most of the information since last update of $\lambda$ is used. A simple rule of thumb is to set $\beta = \gamma = \kappa = \eta = 1 - \sqrt[M]{0.01}$ which means that the weight of the $M^{\text{th}}$ term of the exponentially weighted sum is $0.01$. This worked well in our experiments. Often data streams follow different periodic patterns, thus randomly selecting when to update $\lambda$ can sometimes be useful.

The main challenge with the \textit{HIL} approach is that the quantile estimates and associated MSEs are restarted after an update and must have converged before a new update is performed. This limits how rapidly and smoothly the procedure can adapt to changes in the data stream dynamics. Further since $\lambda$ is updated rarely, a fairly large value of $a$, say $2$, must be used limiting any fine tuning of $\lambda$.

\section{Synthetic Experiments}
\label{sec:synt}

\begin{figure}
  \centering
  \includegraphics[width = \textwidth]{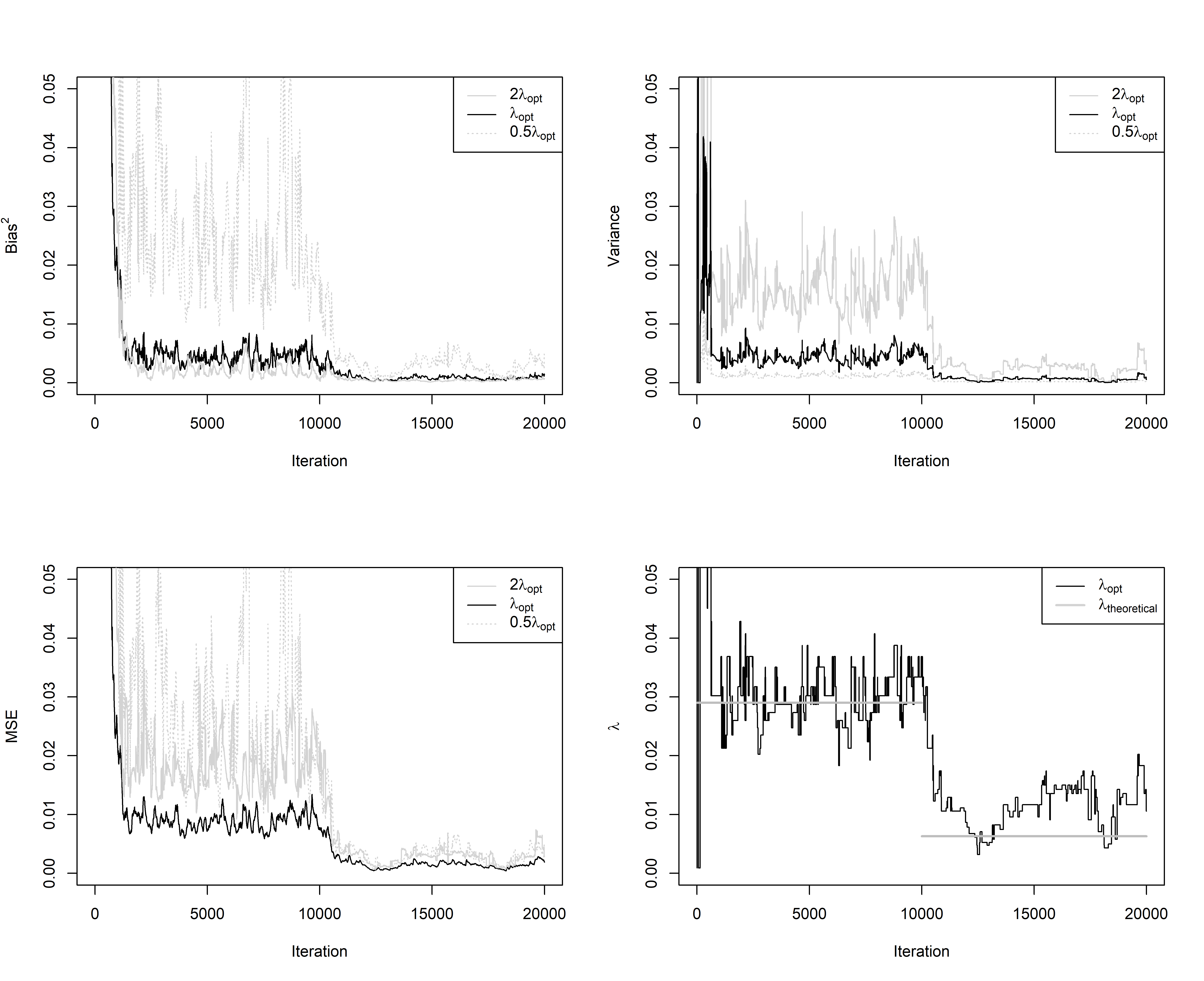}
  \caption{The panels from top left to bottom right shows tracking of Bias$^2$, Variance, MSE and the resulting recursive updating of $\lambda$. The symbol $\lambda{\text{opt}}$ refers to the value resulting in the minimal estimated MSE according to Eq. \eqref{eq:15}. The gray lines in the bottom right panel refers to theoretically optimal values of $\lambda$.}
  \label{fig:2}
\end{figure}

DUMIQE was used for all the experiments in this section. Consider a normally distributed data stream where the expectation changes between slow and rapid dynamics
\begin{align}
  \label{eq:8}
  \begin{split}
  f_t(x) &= N\left(\mu + b \sin\left(\frac{2\pi}{\tau(n)}n\right), \sigma\right) \\[2mm]
  \tau(n) &= \tau_1 I(\text{mod}\, n < T) + \tau_2 I(T \leq \text{mod}\, n < 2T)    
  \end{split}
\end{align}
with $\tau_1 = 500$, $\tau_2 = 10^4$, $\mu = 8$, $b = 2$, $\sigma = 1$, $T = 10^4$.
Figure \ref{fig:2} shows tracking of Bias$^2$, Variance, MSE and values of $\lambda$ using the \textit{Oracle} approach with $\beta = \gamma = \kappa = \eta = 1 - \sqrt[1000]{0.01} = 0.005$. Further, the following values of $\lambda$ were used $\lambda_1 = \exp(-7), \lambda_2 = \exp(-6.95), \ldots, \exp(0)$. 

\begin{figure}
  \centering
  \includegraphics[width = \textwidth]{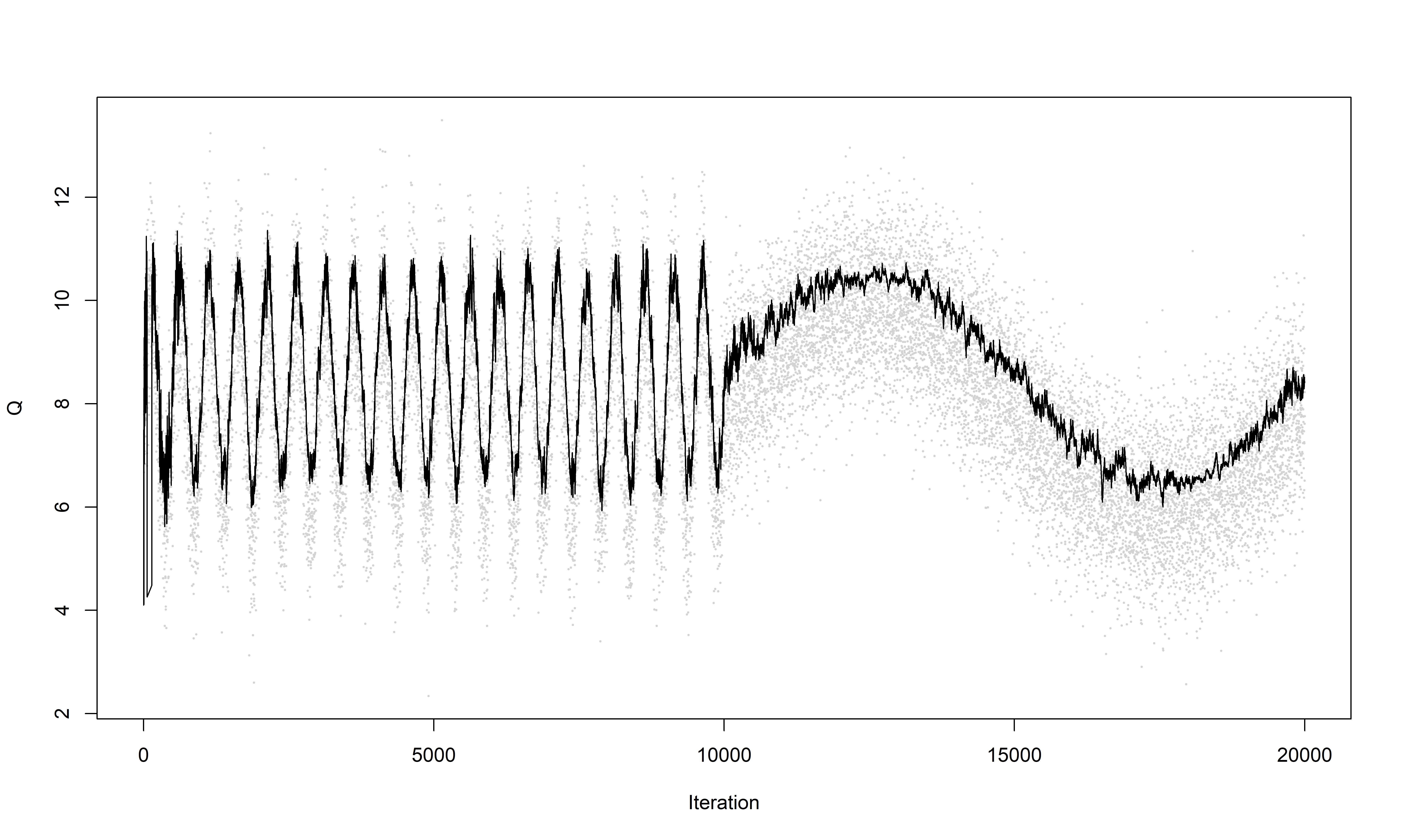}
  \caption{The gray dots represent the data stream in \eqref{eq:8} and the black line is tracking of the $q = 0.7$ quantile with the \textit{Oracle} approach.}
  \label{fig:1}
\end{figure}

Figure \ref{fig:1} shows the resulting tracking.
Top left panel of Figure \ref{fig:2} shows that using a high value of $\lambda$, i.e. $2\lambda{\text{opt}}$, (solid gray curve) results in small bias since the quantile algorithm is able to efficiently track the dynamic changes of the data stream. On the other, as shown in the in the upper right panel, the estimation variance will be high since the step size is high. The MSE in the bottom left panel is further the sum of these to quantities and finding the best $\lambda$ can be interpreted as a continuous battle between tracking bias and variance.
The bottom right panel shows that the procedure quite rapidly finds values centered around the theoretically optimal $\lambda$. When the dynamics of the data stream changes after $10,000$ iterations, the procedure rapidly selects an estimator with a smaller value of $\lambda$ however slightly higher then the theoretical optimal $\lambda$. It is important to recall that Eq. \eqref{eq:15} is only an approximation of the true MSE that is further estimated. We see that both bias and variance are smaller after $10,000$ iterations which is as expected since slow dynamics are easier to track. 

In Figure \ref{fig:1} we see that for the first iterations, the tracking is erratic until a reliable estimates of MSE is obtained. 

\begin{figure}
  \centering
  \includegraphics[width = \textwidth]{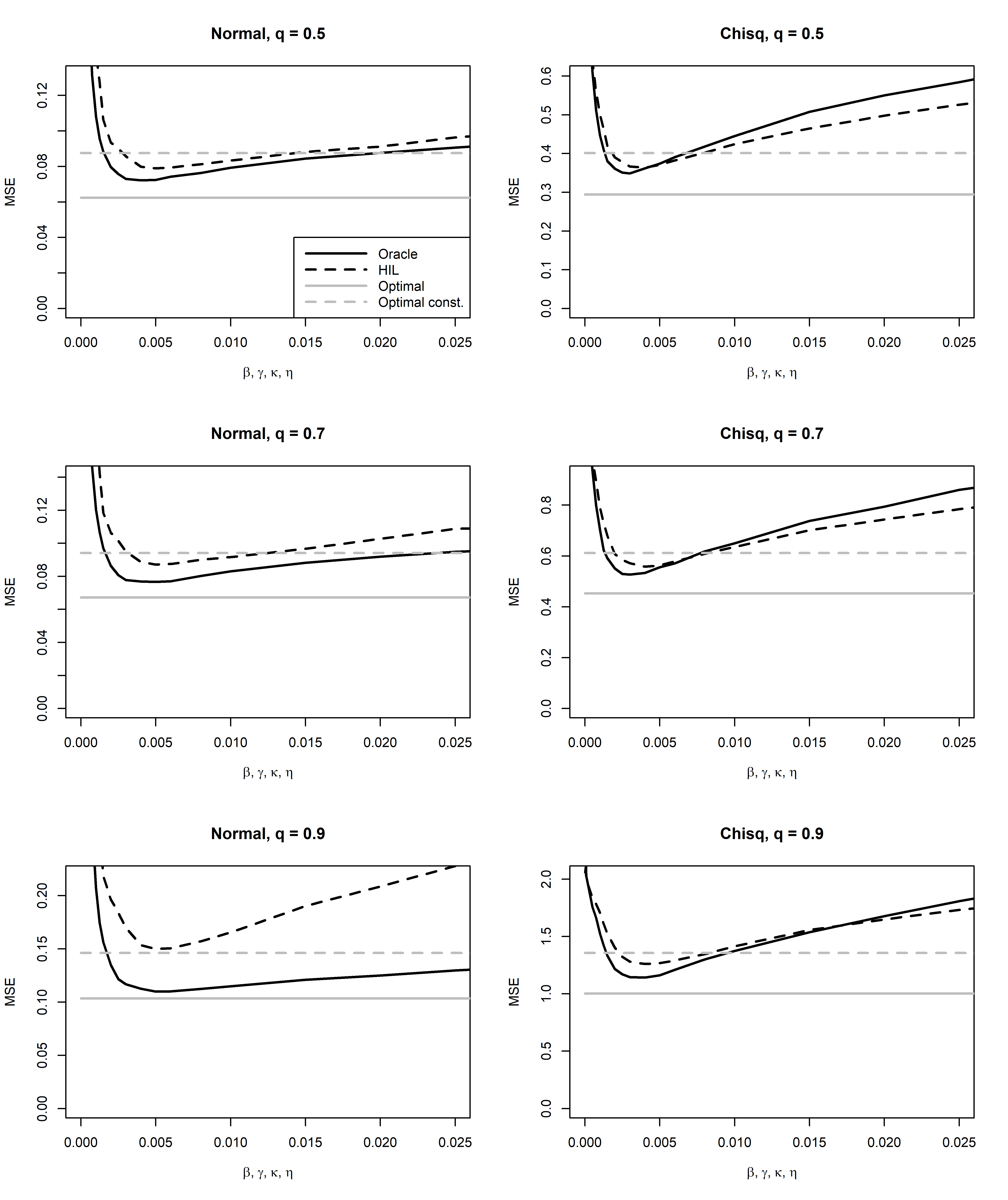}
  \caption{The left and right panels show results for the normal and $\chi^2$ distributed data streams. The rows from top to bottom shows results from tracking of the $q = 0.5, 0.7$ and $0.9$ quantiles, respectively.}
  \label{fig:3}
\end{figure}

We now analyze the performance of the suggested procedure in more detail. Consider again the data stream in \eqref{eq:8} as well as a $\chi^2$ distributed stream
\begin{align}
  \label{eq:9}
  \begin{split}
  f_t(x) &= \chi^2\left(\nu + b \sin\left(\frac{2\pi}{\tau(n)}n\right)\right) \\[2mm]
  \tau(n) &= \tau_1 I(\text{mod}\, n < T) + \tau_2 I(T \leq \text{mod}\, n < 2T)    
  \end{split}
\end{align}
using the same values $\tau_1$, $\tau_2$, $T$ and $b$ as above and $\nu = 6$. $\chi^2(\nu)$ represents the $\chi^2$ distribution with $\nu$ degrees of freedom. The $\chi^2$ distributed stream is challenging since both the expectation and variance change with time. For the \textit{HIL} approach, $a = 1.5$ and $M = 1000 + U$ was used, where $U$ was uniformly distributed on the interval $[0,1000]$, i.e. $\lambda$ was updated on average every 1500 iteration. We tracked the $q = 0.5, 0.7$ and $0.9$ quantiles and used $\widetilde{q} = 0.6, 0.6$ and $0.8$, respectively. 
To remove any Monte Carlo error, data streams were ran for a total of $N = 10^7$ iterations and observed tracking MSE were computed
\begin{equation}
  \label{eq:10}
  \text{MSE}\, = \frac{1}{N} \sum_{t=1}^N \left(\widehat{Q}_{t,q} - Q_{t,q}\right)^2
\end{equation}
The results are shown in Figure \ref{fig:3}.
Let MSE$_{\tau_1}^{\text{theo}}$ and MSE$_{\tau_2}^{\text{theo}}$ represent the theoretically minimum tracking MSE for a data stream with constant fast dynamics ($\tau_1$) and constant slow dynamics ($\tau_2$), respectively. The data streams above, consist of an equal amount of fast and slow dynamics and thus the theoretically minimum tracking error will be $0.5 \, \text{MSE}_{\tau_1}^{\text{theo}} + 0.5 \, \text{MSE}_{\tau_2}^{\text{theo}}$ and is shown as gray solid lines in Figure \ref{fig:3}. We further computed the theoretically minimum tracking MSE for the streams above using a constant value of $\lambda$ and is shown as gray dashed lines. 
The results reveal that the \textit{Oracle} approach is close to the theoretically optimal tracking error for all the cases. The \textit{HIL} approach also performs well, but not as good as \textit{Oracle} approach, which is as expected given the discussions at the end of Section \ref{sec:MSE}. Optimal values of the tuning parameters in the procedures ($\beta, \gamma, \kappa, \eta$) are close to the suggested rule of thumb $1 - \sqrt[M]{0.01} = 0.0031$. 

\section{Real-Life Data Example - Tracking of Twitter Traffic} 
\label{sec:real}

In this section we present a real-world example for benchmarking the procedures. Analysis of Twitter steaming data is an important application in research but also in practice.  Efficient analysis of Twitter streams allows to for example perform detection of natural disasters or is used to predict election outcomes. 

For our experiment, let $T_t$ represent the time point when tweet number $n$ is received from a stream of tweets. In this section we will track quantiles of the quantity $R_t = (T_t - T_{t-1})^{-1}$ which can be interpreted as the frequency of received tweets. The quantity can for instance be used to real world event detection in the sense that if the number of received tweets is increasing, $R_t$ will increase \citep{atefeh2015survey,hasan2017survey}. 

We consider a dataset consisting of the time points of every posted tweet by Norwegian users before and after the horrific terrorist attack July 22 2011 \citep{sollid2012oslo}. The time stamp for each tweet were given as whole second. To reconstruct a representation of the true time stamps a uniformly drawn value between zero and one second were added to each time stamp. The data stream is massive and we assumed no knowledge about suitable values for $\lambda$ and thus used the \textit{HIL} approach. We tracked quantiles of $R_t$ using the Frugal estimator \citep{ma2013frugal}. Since the update size of the Frugal estimator is independent of the scale of the data, the suggested procedure must adjust $\lambda$ if the scale of the $R_t$ changes as well as if the dynamics changes. The value of $\lambda$ was adjusted after every $M = 1000$ tweet received, which on average was about every 18$^{\text{th}}$ minute in real-time. The value of $\lambda$ was scaled by $a = 1.5$ and the tuning parameters were $\beta, \gamma, \kappa$ and $\eta$ was set to $0.005$ in accordance with the rule of thumb in Section \ref{sec:MSE}. 

\begin{figure}
  \centering
  \includegraphics[width = \textwidth]{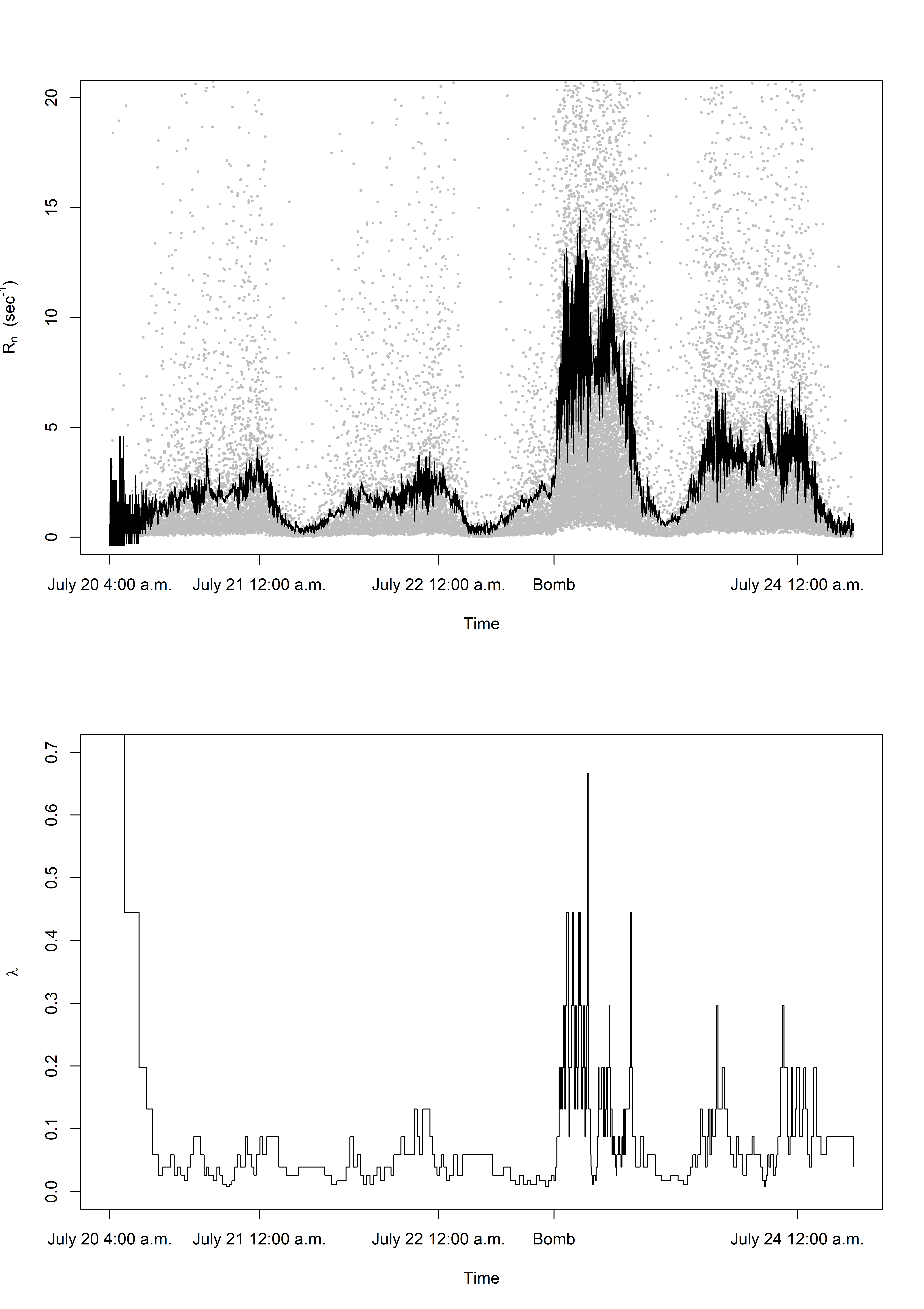}
  \caption{Twitter data example. Upper panel: Gray dots show the observed data stream $R_t$ (every 10$^{\text{th}}$ shown) and the black curve tracking of the $q = 0.7$ quantile using the Frugal estimator. Bottom panel: the values of $\lambda$.}
  \label{fig:4}
\end{figure}

The results are shown in Figure \ref{fig:4}. The attack was initiated by a bomb going of in Oslo July 22 at 3:25 p.m. local time and is marked on the $x$ axis. The black curve in the upper panel shows the tracking of the $q = 0.7$ quantile of $R_t$. The gray dots show the observed $R_t$. The bottom panel shows the value of $\lambda$ in every iteration. The Frugal estimator was initiated with $\lambda = 1$ and the procedure rapidly adjusted it to a more suitable value.
During the terrorist attack at some point a bomb placed by the terrorist exploded which had an impact on the tweets.

For the period until the bomb went of it is a trend that $\lambda$ is adjusted to lower values during night time which makes sense since both scale and dynamics of $R_t$ are small. When the bomb went off, $R_t$ changed rapidly. The value $\lambda$ was rapidly increased to be able to efficiently track under the increased scale and dynamics. However, when the data stream stabilizes on a higher scale a few hours of the bomb went of the value of $\lambda$ is slightly decreased. The value of $\lambda$ is again increased when $R_t$ rapidly decreases (rapid dynamics) during the evening of July 22. More or less the same trend in $\lambda$ is observed during July 23, a high value of $\lambda$ is used in the morning to track the increased scale and dynamics, reduction of $\lambda$ during daytime since the dynamics stabilized and again an increase in the value of $\lambda$ in the evening of July 23. Overall the experiment demonstrates that the suggested procedure is able to adjust $\lambda$ to obtain efficient quantile tracking in all parts of the complex data stream.

\section{Closing Remarks}
\label{sec:closrem}

Surprisingly little attention has been given to automatic adjustment of the values of tuning parameters of incremental quantile estimators. In this paper we develop two simple procedures to achieve this. Both procedures are based on estimating the current tracking MSE and use this to efficiently track the true quantiles. The \textit{Oracle} approach tracks the quantile and associated MSE for a wide range of values of the tuning parameter and in each iteration select the quantile estimate with minimal estimated MSE. The second approach only tracks the quantile for three values of the tuning parameter and repeatedly forget the estimate with the highest estimated MSE and add a quantile estimator for another value of the tuning parameter. Both methods are computationally and memory efficient since only a limited set of quantities needs to be computed and stored in each iteration.

The results show that the methods are highly efficient in adjusting the value of $\lambda$ to achieve efficient tracking. The synthetic experiments showed that the resulting tracking error is close to the theoretical minimum. The \textit{Oracle} performs the best, but at a higher computational cost. The real-life data example demonstrated that the procedure is able to handle complex and massive real-life data streams.

\bibliographystyle{dcu}
\bibliography{bibl}

\end{document}